\documentclass[reprint,amsmath,amssymb,prb,groupedaddress,nofootinbib,twocolumn,superscriptaddress, floatfix]{revtex4-2}

\usepackage{graphicx}
\usepackage{dcolumn}
\usepackage{bm}
\usepackage{acronym}
\usepackage{physics}

\def\[#1\]{\begin{equation}#1\end{equation}}

\newcommand{\tex}[1]{\expval{#1}} 
\newcommand{\Z}{\mathcal{Z}} 
\newcommand{\T}{\mathcal{T}}
\newcommand{\A}{\mathcal{A}}
\newcommand{\p}{\mathcal{P}}

\usepackage{orcidlink}

\usepackage{listings}
\usepackage{xcolor}

\definecolor{codegreen}{rgb}{0,0.6,0}
\definecolor{codegray}{rgb}{0.5,0.5,0.5}
\definecolor{codepurple}{rgb}{0.58,0,0.82}
\definecolor{backcolour}{rgb}{0.95,0.95,0.92}
\definecolor{urlblue}{HTML}{007bff}

\lstdefinestyle{mystyle}{
    backgroundcolor=\color{backcolour},   
    commentstyle=\color{codegreen},
    keywordstyle=\color{magenta},
    numberstyle=\tiny\color{codegray},
    stringstyle=\color{codepurple},
    basicstyle=\ttfamily\footnotesize,
    breakatwhitespace=false,         
    breaklines=true,                 
    captionpos=b,                    
    keepspaces=true,                 
    numbers=left,                    
    numbersep=5pt,                  
    showspaces=false,                
    showstringspaces=false,
    showtabs=false,                  
    tabsize=2
}

    \newacro{FK}{Falikov-Kimball}
    \newacro{CDW}[CDW]{charge-density wave}
    \newacro{MCMC}{Markov chain Monte Carlo}
    \newacro{ED}{exact diagonalisation}
    \newacro{TMM}{Transfer Matrix Methods}
    \newacro{IPR}{Inverse Participation Ratio}
    \newacro{DOS}{density of states}
    \newacro{FTPT}{finite temperature phase transition}
    \newacro{LRI}{Long-Range Ising}

\usepackage{hyperref}
\hypersetup{
    colorlinks = true,
    linkcolor  = urlblue,
    citecolor  = urlblue,
    urlcolor   = urlblue,
}

\begin{document}


\title{The one-dimensional Long-Range Falikov-Kimball Model: \\ 
Thermal Phase Transition and Disorder-Free Localisation}

\author{T. Hodson \orcidlink{0000-0002-4121-4772}}
\affiliation{\small Blackett Laboratory, Imperial College London, London SW7 2AZ, United Kingdom}

\author{J. Willsher \orcidlink{0000-0002-6895-6039}}
\affiliation{Department of Physics TQM, Technische Universit{\"a}t M{\"u}nchen, James-Franck-Stra{\ss}e 1, D-85748 Garching, Germany}

\author{J. Knolle \orcidlink{0000-0002-0956-2419}}
\affiliation{Department of Physics TQM, Technische Universit{\"a}t M{\"u}nchen, James-Franck-Stra{\ss}e 1, D-85748 Garching, Germany}
\affiliation{Munich Center for Quantum Science and Technology (MCQST), 80799 Munich, Germany}
\affiliation{\small Blackett Laboratory, Imperial College London, London SW7 2AZ, United Kingdom}
\date{\today}

\begin{abstract}

 Disorder or interactions can turn metals into insulators. One of the simplest settings to study this physics is given by the \ac{FK} model, which describes itinerant fermions interacting with a classical Ising background field. Despite the translational invariance of the model, inhomogenous configurations of the background field give rise to effective disorder physics which lead to a rich phase diagram in two (or more) dimensions with finite temperature charge density wave (CDW) transitions and interaction-tuned Anderson versus Mott localized phases. Here, we propose a generalised \ac{FK} model in one dimension with long-range interactions which shows a similarly rich phase diagram. We use an exact Markov Chain Monte Carlo method to map the phase diagram and compute the energy resolved localisation properties of the fermions. We compare the behaviour of this transitionally invariant model to an Anderson model of uncorrelated binary disorder about a background CDW field which confirms that the fermionic sector only fully localizes for very large system sizes.
\end{abstract}

\maketitle

\begin{figure}[!ht]
  \centering
    \includegraphics[width=\columnwidth]{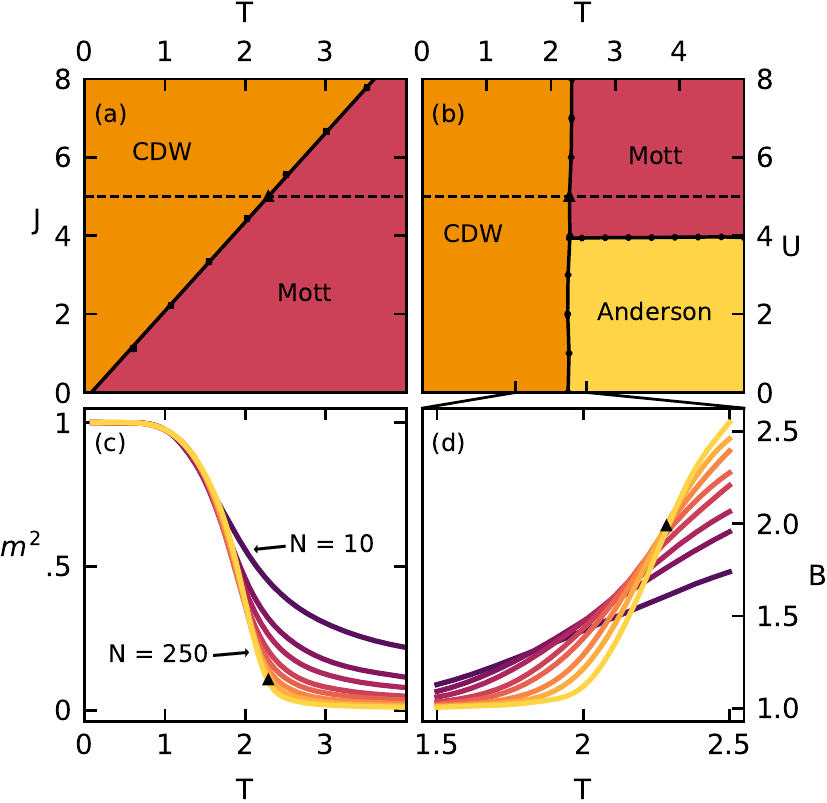}
  \caption{\label{fig:phase_diagram} Phase diagrams of the long-range 1D \ac{FK} model. (a) The TJ plane at \(U = 5\): the \ac{CDW} ordered phase is separated from a disordered Mott insulating (MI) phase by a critical temperature \(T_c\), linear in J. (b) The TU plane at \(J = 5\): the disordered phase is split into two: at large/small U there's a MI/Anderson phase characterised by the presence/absence of a gap at \(E=0\) in the single particle energy spectrum. \(U_c\) is independent of temperature. (c) The order parameter \(\tex{m^2}\) describing the onset of the \ac{CDW} phase of the long-range 1D \ac{FK} model at low temperature with staggered magnetisation \(m = N^{-1} \sum_i (-1)^i n_i\). (d) The crossing of the Binder cumulant, \(B = \tex{m^4} / \tex{m^2}^2\), with system size provides a diagnostic that the phase transition is not a finite size effect, it's used to estimate the critical lines shown in (a) and (b). All plots use system sizes \(N = [10,20,30,50,70,110,160,250]\) and parameter values \(U = 5,\;J = 5,\;\alpha = 1.25\) except where explicitly varied.}
\end{figure}

\section{Introduction}
The \acl{FK} model is one of the simplest models of the correlated electron problem. It captures the essence of the interaction between itinerant and localized electrons, equivalent to a model of hopping fermions coupled to a classical Ising field. It was originally introduced to explain the metal-insulator transition in f-electron systems but in its long history it has been interpreted variously as a model of electrons and ions, binary alloys or of crystal formation~\cite{hubbardj.ElectronCorrelationsNarrow1963, falicovSimpleModelSemiconductorMetal1969,gruberFalicovKimballModelReview1996,gruberFalicovKimballModel2006}. Despite its simplicity, the \ac{FK} model has a rich phase diagram in \(D \geq 2\) dimensions. For example, it shows an interaction-induced gap opening even at high temperatures, similar to the corresponding Hubbard Model~\cite{brandtThermodynamicsCorrelationFunctions1989}. Moreover, it has been a test-bed for many-body methods, interest took off when an exact DMFT solution in the infinite dimensional case was found~\cite{antipovCriticalExponentsStrongly2014,ribicNonlocalCorrelationsSpectral2016,freericksExactDynamicalMeanfield2003,herrmannNonequilibriumDynamicalCluster2016}. 

The presence of the classical field makes the model amenable to an exact numerical treatment at finite temperature via a sign problem free \ac{MCMC} algorithm~\cite{devriesGapsDensitiesStates1993,devriesSimplifiedHubbardModel1993,antipovInteractionTunedAndersonMott2016,debskiPossibilityDetectionFinite2016,herrmannSpreadingCorrelationsFalicovKimball2018,maskaThermodynamicsTwodimensionalFalicovKimball2006}. The \ac{MCMC} treatment motivates a view of the classical background field as a disorder potential, which suggests an intimate link to localisation physics. Indeed, thermal fluctuations of the classical sector act as disorder potentials drawn from a thermal distribution and the emergence of disorder in a translationally invariant Hamiltonian links the \ac{FK} model to recent interest in disorder-free localisation~\cite{smithDisorderFreeLocalization2017,smithDynamicalLocalizationMathbbZ2018,brenesManyBodyLocalizationDynamics2018}.

Dimensionality is crucial for the physics of both localisation and \acp{FTPT}. In 1D, disorder generally dominates, even the weakest disorder exponentially localises \textit{all} single particle eigenstates. Only longer-range correlations of the disorder potential can potentially induce delocalization~\cite{aubryAnalyticityBreakingAnderson1980,dassarmaLocalizationMobilityEdges1990,dunlapAbsenceLocalizationRandomdimer1990}. Thermodynamically, short-range interactions cannot overcome thermal defects in 1D which prevents ordered phases at nonzero temperature~\cite{andersonAbsenceDiffusionCertain1958,goldshteinPurePointSpectrum1977a,abrahamsScalingTheoryLocalization1979,kramerLocalizationTheoryExperiment1993}. However, the absence of an \ac{FTPT} in the short ranged \ac{FK} chain is far from obvious because the Ruderman-Kittel-Kasuya-Yosida (RKKY) interaction mediated by the fermions~\cite{kasuyaTheoryMetallicFerro1956,rudermanIndirectExchangeCoupling1954,vanvleckNoteInteractionsSpins1962,yosidaMagneticPropertiesCuMn1957} decays as \(r^{-1}\) in 1D~\cite{rusinCalculationRKKYRange2017a}. This could in principle induce the necessary long-range interactions for the classical Ising background~\cite{thoulessLongRangeOrderOneDimensional1969,peierlsIsingModelFerromagnetism1936}. However, Kennedy and Lieb established rigorously that at half-filling a \ac{CDW} phase only exists at \(T = 0\) for the 1D \ac{FK} model~\cite{kennedyItinerantElectronModel1986}. 

Here, we construct a generalised one-dimensional \ac{FK} model with long-range interactions which induces the otherwise forbidden \ac{CDW} phase at non-zero temperature. We find a rich phase diagram with a CDW FTPT and interaction-tuned Anderson versus Mott localized phases similar to the 2D \ac{FK} model~\cite{antipovInteractionTunedAndersonMott2016}. We explore the localization properties of the fermionic sector and find that the localisation lengths vary dramatically across the phases and for different energies. Although moderate system sizes indicate the coexistence of localized and delocalized states within the CDW phase, we find quantitatively similar behaviour in a model of uncorrelated binary disorder on a \ac{CDW} background. For large system sizes, i.e. for our 1D disorder model we can treat linear sizes of several thousand sites, we find that all states are eventually localized with a localization length which diverges towards zero temperature.   

\section{The Long-Ranged Falikov-Kimball Model}
We interpret the \ac{FK} model as a model of spinless fermions, \(c^\dag_{i}\), hopping on a 1D lattice against a classical Ising spin background, \(S_i \in {\pm \frac{1}{2}}\). The fermions couple to the spins via an onsite interaction with strength \(U\) which we supplement by a long-range interaction, \(J_{ij} = 4\kappa J (-1)^{\abs{i-j}} \abs{i-j}^{-\alpha}\), between the spins. The normalisation, \(\kappa^{-1} = \sum_{i=1}^{N} i^{-\alpha}\), renders the 0th order mean field critical temperature independent of system size. The hopping strength of the electrons, \(t = 1\), sets the overall energy scale and we concentrate throughout on the particle-hole symmetric point at zero chemical potential and half filling~\cite{gruberFalicovKimballModelReview1996}.
~
\begin{align}
H_{\mathrm{FK}} = & \;U \sum_{i} S_i\;(c^\dag_{i}c_{i} - \tfrac{1}{2}) -\;t \sum_{i} c^\dag_{i}c_{i+1} + c^\dag_{i+1}c_{i}\\ 
 &  + \sum_{i, j}^{N} J_{ij}  S_i S_j \nonumber
\label{eq:HFK}
\end{align}

In two or more dimensions, the \(J\!=0\!\) \ac{FK} model has a \ac{FTPT} to the \ac{CDW} phase with non-zero staggered magnetisation \(m = N^{-1} \sum_i (-1)^i \; S_i\)~\cite{antipovInteractionTunedAndersonMott2016,maskaThermodynamicsTwodimensionalFalicovKimball2006}. This only exists at zero temperature in the short ranged 1D model~\cite{kennedyItinerantElectronModel1986}. To study the \ac{CDW} phase at finite temperature in 1D, we add an additional coupling that is both long-ranged and staggered by a factor \((-1)^{|i-j|}\). The additional coupling stabilises the Antiferromagnetic (AFM) order of the Ising spins which promotes the finite temperature \ac{CDW} phase of the fermionic sector.

Taking the limit \(U = 0\) decouples the spins from the fermions, which gives a spin sector governed by a classical \ac{LRI} model. Note, the transformation of the spins \(S_i \to (-1)^{i} S_i\) maps the AFM model to the FM one. We recall that Peierls' classic argument can be extended to show that, for the 1D \ac{LRI} model, a power law decay of \(\alpha < 2\) is required for a \ac{FTPT} as the energy of defect domain then scales with the system size and can overcome the entropic contribution. A renormalisation group analysis supports this finding and shows that the critical exponents are only universal for \(\alpha \leq 3/2\)~\cite{ruelleStatisticalMechanicsOnedimensional1968,thoulessLongRangeOrderOneDimensional1969,angeliniRelationsShortrangeLongrange2014}. In the following, we choose \(\alpha = 5/4\) to avoid this additional complexity. 

To improve the scaling of finite size effects, we make the replacement \( \abs{i - j}^{-\alpha} \rightarrow \abs{f(i - j)}^{-\alpha}\), in both \(J_{ij}\) and \(\kappa\), where \(f(x) = \frac{N}{\pi}\sin \frac{\pi x}{N}\), which is smooth across the circular boundary~\cite{fukuiOrderNClusterMonte2009}. We only consider even system sizes given that odd system sizes are not commensurate with a \ac{CDW} state.

\section{The Phase Diagram} 
Figs.~[\ref{fig:phase_diagram}a] and [\ref{fig:phase_diagram}b] show the phase diagram for constant \(U=5\) and constant \(J=5\), respectively. We determined the transition temperatures from the crossings of the Binder cumulants \(B_4 = \tex{m^4}/\tex{m^2}^2\)~\cite{binderFiniteSizeScaling1981}. For a representative set of parameters, Fig.~[\ref{fig:phase_diagram}c] shows the order parameter \(\tex{m}^2\). Fig.~[\ref{fig:phase_diagram}d] shows the Binder cumulants, both as functions of system size and temperature. The crossings confirm that the system has a \ac{FTPT} and that the ordered phase is not a finite size effect. 

The CDW transition temperature is largely independent from the strength of the interaction \(U\). This demonstrates that the phase transition is driven by the long-range term \(J\) with little effect from the coupling to the fermions \(U\).  The physics of the spin sector in our long-range \ac{FK} model mimics that of the \ac{LRI} model and is not significantly altered by the presence of the fermions, which shows that the long range tail expected from a basic fermion mediated RKKY interaction between the Ising spins is absent. 

Our main interest concerns the additional structure of the fermionic sector in the high temperature phase. Following Ref.~\cite{antipovInteractionTunedAndersonMott2016}, we can distinguish between the Mott and Anderson insulating phases. The former is characterised by a gapped \ac{DOS} in the absence of a \ac{CDW}. Thus, the opening of a gap for large \(U\) is distinct from the gap-opening induced by the translational symmetry breaking in the CDW state below \(T_c\), see also Fig.~[\ref{fig:band_opening}a]. The Anderson phase is gapless but, as we explain below, shows localised fermionic eigenstates. 

\section{Markov Chain Monte Carlo and Emergent Disorder}
The results for the phase diagram were obtained with a classical \ac{MCMC} method which we discuss in the following. It allows us to solve our long-range \ac{FK} model efficiently, yielding unbiased estimates of thermal expectation values and linking it to disorder physics in a translationally invariant setting.

Since the spin configurations are classical, the Hamiltonian can be split into a classical spin part \(H_s\) and an operator valued part \(H_c\).
\begin{align}
H_s& = - \frac{U}{2}S_i + \sum_{i, j}^{N} J_{ij} S_i S_j \\
H_c& = \sum_i U S_i c^\dag_{i}c_{i} -t(c^\dag_{i}c_{i+1} + c^\dag_{i+1}c_{i}) 
\end{align}
The partition function can then be written as a sum over spin configurations, \(\vec{S} = (S_0, S_1...S_{N-1})\):
\begin{align}
\Z = \Tr e^{-\beta H}= \sum_{\vec{S}} e^{-\beta H_s} \Tr_c e^{-\beta H_c} .
\end{align}
The contribution of \(H_c\) to the grand canonical partition function can be obtained by performing the sum over eigenstate occupation numbers giving \(-\beta F_c[\vec{S}] = \sum_k \ln{(1 + e^{- \beta \epsilon_k})}\) where \({\epsilon_k[\vec{S}]}\) are the eigenvalues of the matrix representation of \(H_c\) determined through exact diagonalisation. This gives a partition function containing a classical energy which corresponds to the long-range interaction of the spins, and a free energy which corresponds to the quantum subsystem. 
\begin{align}
\Z = \sum_{\vec{S}} e^{-\beta H_S[\vec{S}] - \beta F_c[\vec{S}]} = \sum_{\vec{S}} e^{-\beta E[\vec{S}]}
\end{align}

\ac{MCMC} defines a weighted random walk over the spin states \((\vec{S}_0, \vec{S}_1, \vec{S}_2, ...)\), such that the likelihood of visiting a particular state converges to its Boltzmann probability \(p(\vec{S}) = \Z^{-1} e^{-\beta E}\)~\cite{binderGuidePracticalWork1988,kerteszAdvancesComputerSimulation1998,wolffMonteCarloErrors2004}. Hence, any observable can be estimated as a mean over the states visited by the walk.
\begin{align} \label{eq:thermal_expectation}
\tex{O}& = \sum_{\vec{S}} p(\vec{S}) \tex{O}_{\vec{S}} = \sum_{i = 0}^{M} \tex{O}_{\vec{S}_i} + \mathcal{O}(\tfrac{1}{\sqrt{M}})\\
\tex{O}_{\vec{S}}& = \sum_{\nu} n_F(\epsilon_{\nu}) \expval{O}{\nu}
\end{align}
Where \(\nu\) runs over the eigenstates of \(H_c\) for a particular spin configuration and \(n_F(\epsilon) = \left(e^{-\beta\epsilon} + 1\right)^{-1}\) is the Fermi function.

The choice of the transition function for \ac{MCMC} is under-determined as one only needs to satisfy a set of balance conditions for which there are many solutions~\cite{kellyReversibilityStochasticNetworks1981}. Here, we incorporate a modification to the standard Metropolis-Hastings algorithm~\cite{hastingsMonteCarloSampling1970} gleaned from Krauth~\cite{krauthIntroductionMonteCarlo1998}. Let us first recall the standard algorithm which decomposes the transition probability into \(\T(a \to b) = \p(a \to b)\A(a \to b)\). Here,  \(\p\) is the proposal distribution that we can directly sample from while \(\A\) is the acceptance probability. The standard Metropolis-Hastings choice is 
\[\A(a \to b) = \min\left(1, \frac{\p(b\to a)}{\p(a\to b)} e^{-\beta \Delta E)}\right)\;,\]
with \(\Delta E = E_b - E_a\). The walk then proceeds by sampling a state \(b\) from \(\p\) and moving to \(b\) with probability \(\A(a \to b)\). The latter operation is typically implemented by performing a transition if a uniform random sample from the unit interval is less than \(\A(a \to b)\) and otherwise repeating the current state as the next step in the random walk. The proposal distribution is often symmetric so does not appear in \(\A\). Here, we flip a small number of sites in \(b\) at random to generate proposals, which is indeed symmetric. 

In our computations~\cite{hodsonMCMCFKModel2021} we employ a modification of the algorithm which is based on the observation that the free energy of the \ac{FK} system is composed of a classical part which is much quicker to compute than the quantum part. Hence, we can obtain a computational speedup by first considering the value of the classical energy difference \(\Delta H_s\) and rejecting the transition if the former is too high. We only compute the quantum energy difference \(\Delta F_c\) if the transition is accepted. We then 
perform a second rejection sampling step based upon it. This corresponds to two nested comparisons with the majority of the work only occurring if the first test passes and has the acceptance function 
\[\A(a \to b) = \min\left(1, e^{-\beta \Delta H_s)}\right)\min\left(1, e^{-\beta \Delta F_c)}\right)\;.\]

See Appendix~\ref{app:balance} for a proof that this satisfies the detailed balance condition.

For the model parameters used in Fig.~\ref{fig:indiv_IPR}, we find that with our new scheme the matrix diagonalisation is skipped around 30\% of the time at \(T = 2.5\) and up to 80\% at \(T = 1.5\). We observe that for  \(N = 50\), the matrix diagonalisation, if it occurs, occupies around 60\% of the total computation time for a single step. This rises to 90\% at N = 300 and further increases for larger N. We therefore get the greatest speedup for large system sizes at low temperature where many prospective transitions are rejected at the classical stage and the matrix computation takes up the greatest fraction of the total computation time. The upshot is that we find a  speedup of up to a factor of 10 at the cost of very little extra algorithmic complexity. 

Our two-step method should be distinguished from the more common method for speeding up \ac{MCMC} which is to add asymmetry to the proposal distribution to make it as similar as possible to \(\min\left(1, e^{-\beta \Delta E}\right)\). This reduces the number of rejected states, which brings the algorithm closer in efficiency to a direct sampling method. However it comes at the expense of requiring a way to directly sample from this complex distribution, a problem which \ac{MCMC} was employed to solve in the first place. For example, recent work trains restricted Boltzmann machines (RBMs) to generate samples for the proposal distribution of the \ac{FK} model~\cite{huangAcceleratedMonteCarlo2017}. The RBMs are chosen as a parametrisation of the proposal distribution that can be efficiently sampled from while offering sufficient flexibility that they can be adjusted to match the target distribution. Our proposed method is considerably simpler and does not require training while still reaping some of the benefits of reduced computation.

\section{Localisation Properties}

The \ac{MCMC} formulation suggests viewing the spin configurations as a form of annealed binary disorder whose probability distribution is given by the Boltzmann weight \(e^{-\beta H_S[\vec{S}] - \beta F_c[\vec{S}]}\). This makes apparent the link to the study of disordered systems and Anderson localisation. While these systems are typically studied by defining the probability distribution for the quenched disorder potential externally, here we have a translation invariant system with disorder as a natural consequence of the Ising background field conserved under the dynamics.  

In the limits of zero and infinite temperature, our model becomes a simple tight-binding model for the fermions. At zero temperature, the spin background is in one of the two translation invariant AFM ground states with two gapped fermionic CDW bands at energies
\[E_{\pm} = \pm\sqrt{\frac{1}{4}U^2 + 2t^2(1 + \cos ka)^2}\;.\]

At infinite temperature, all the spin configurations become equally likely and the fermionic model reduces to one of binary uncorrelated disorder in which all eigenstates are Anderson localised~\cite{abrahamsScalingTheoryLocalization1979}. An Anderson localised state centered around \(r_0\) has magnitude that drops exponentially over some localisation length \(\xi\) i.e \(|\psi(r)|^2 \sim \exp{-\abs{r - r_0}/\xi}\). Calculating $\xi$ directly is numerically demanding. Therefore, we determine if a given state is localised via the energy-resolved \ac{IPR} and the \ac{DOS} defined as
\begin{align}
\mathrm{DOS}(\vec{S}, \omega)& = N^{-1} \sum_{i} \delta(\epsilon_i - \omega)\\
\mathrm{IPR}(\vec{S}, \omega)& = \; N^{-1} \mathrm{DOS}(\vec{S}, \omega)^{-1} \sum_{i,j} \delta(\epsilon_i - \omega)\;\psi^{4}_{i,j}
\end{align}
where \(\epsilon_i\) and \(\psi_{i,j}\) are the \(i\)th energy level and \(j\)th element of the corresponding eigenfunction, both dependent on the background spin configuration \(\vec{S}\). 

The scaling of the IPR with system size
\[\mathrm{IPR} \propto N^{-\tau}\]
depends on the localisation properties of states at that energy. For delocalised states, e.g. Bloch waves, \(\tau\) is the physical dimension. For fully localised states \(\tau\) goes to zero in the thermodynamic limit. However, for special types of disorder such as binary disorder, the localisation lengths can be large comparable to the system size at hand, which can make it difficult to extract the correct scaling. An additional complication arises from the fact that the scaling exponent may display intermediate behaviours for correlated disorder and in the vicinity of a localisation-delocalisation transition~\cite{kramerLocalizationTheoryExperiment1993, eversAndersonTransitions2008a}. The thermal defects of the CDW phase lead to a binary disorder potential with a finite correlation length, which in principle could result in delocalized eigenstates.  

The key question for our system is then: How is the \(T=0\) CDW phase with fully delocalized fermionic states connected to the fully localized phase at high temperatures? 
%
%
\begin{figure}
  \centering
    \includegraphics[width=\columnwidth]{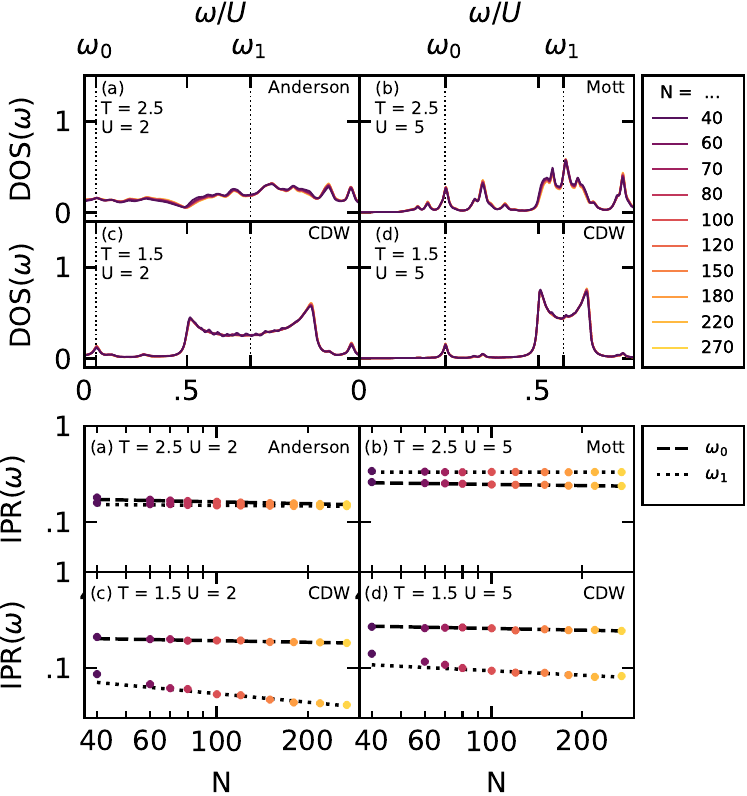}
  \caption{\label{fig:indiv_IPR} Energy resolved DOS(\(\omega\)) and \(\tau\) (the scaling exponent of IPR(\(\omega\)) against system size \(N\)). The left column shows the Anderson phase \(U = 2\) at high \(T = 2.5\) and the CDW phase at low \(T = 1.5\) temperature. IPRs are evaluated for one of the in-gap states \(\omega_0/U = 0.057\) and the center of the band \(\omega_1\) \(U = 0.81\). The right column shows instead the Mott and CDW phases at \(U = 5\) with \(\omega_0/U = 0.24\) and \(\omega_1/U = 0.571\).
  For all the plots \(J = 5,\;\alpha = 1.25\) and the fits for \(\tau\) use system sizes greater than 60. The measured \(\tau_0,\tau_1\) for each figure are: (a) \((0.06\pm0.01, 0.02\pm0.01\) (b) \(0.04\pm0.02, 0.00\pm0.01\) (c) \(0.05\pm0.03, 0.30\pm0.03\)
  (d) \(0.06\pm0.04,  0.15\pm0.05\)
  We show later that the apparent scaling of the IPR
  with system size can be explained by the changing defect density with system size rather than due to delocalisation of the states.}
\end{figure}

\begin{figure}
  \centering
    \includegraphics[width=\columnwidth]{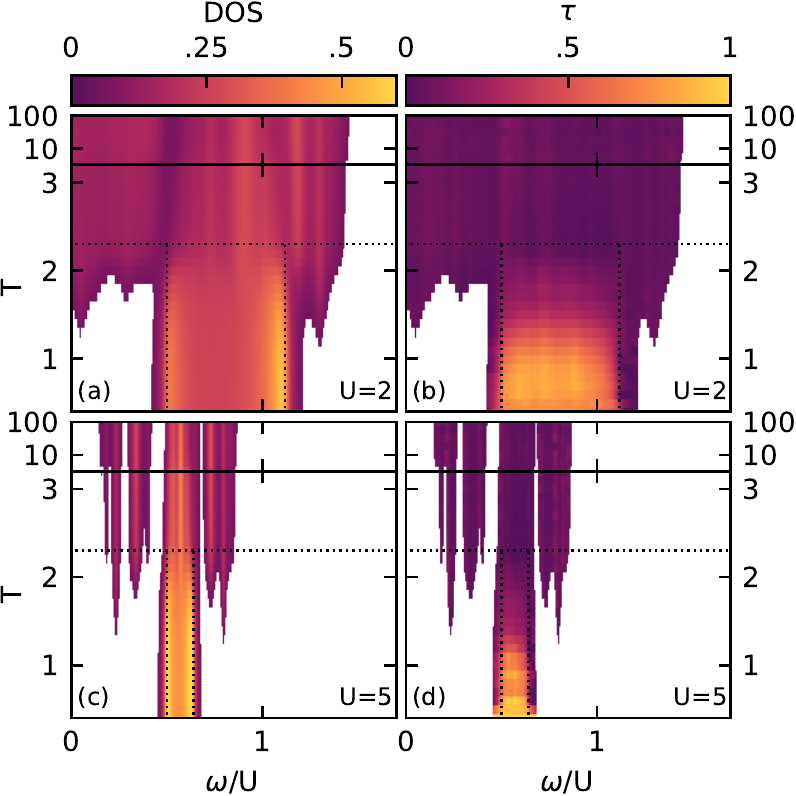}
  \caption{  \label{fig:band_opening} The \acl{DOS} (a and c) and scaling exponent \(\tau\) (b and d) as a function of energy and temperature. (a) and (b) show the system transitioning from the CDW phase to the gapless Anderson insulating one at \(U=2\) while (c) and (d) show the CDW to gapped Mott phase transition at \(U=5\). Regions where the DOS is close to zero are shown a white. The scaling exponent \(\tau\) is obtained from fits to \(IPR(N) = A N^{-\lambda}\) for a range of system sizes.  \(U = 5,\;J = 5,\;\alpha = 1.25\)}
\end{figure}

\begin{figure}[ht]
  \centering
    \includegraphics[width=\columnwidth]{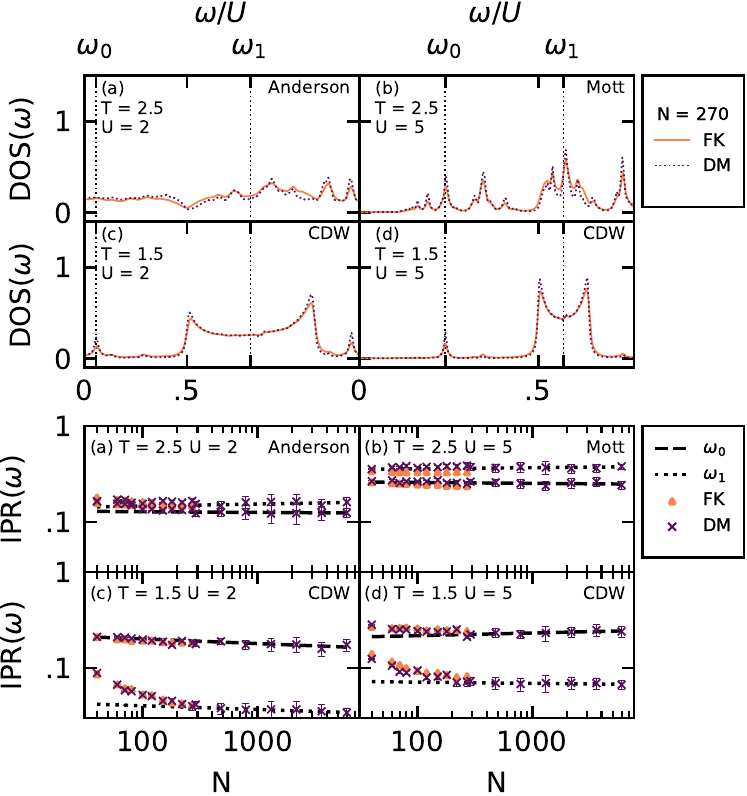}
  \caption{  \label{fig:indiv_IPR_disorder} A comparison of the full \ac{FK} model to a simple binary disorder model (DM) with a CDW wave background perturbed by uncorrelated defects at density \(0 < \rho < 1\) matched to the largest corresponding FK model. As in Fig~\ref{fig:indiv_IPR}, the Energy resolved DOS(\(\omega\)) and \(\tau\) are shown. The DOSs match well and this data makes clear that the apparent scaling of IPR with system size is a finite size effect due to weak localisation~\cite{antipovInteractionTunedAndersonMott2016}, hence all the states are indeed localised as one would expect in 1D. The disorder model \(\tau_0,\tau_1\) for each figure are: (a) \(0.01\pm0.05, -0.02\pm0.06\) (b) \(0.01\pm0.04, -0.01\pm0.04\) (c) \(0.05\pm0.06, 0.04\pm0.06\) (d) \(-0.03\pm0.06, 0.01\pm0.06\). The lines are fit on system sizes \(N > 400\)
  }
\end{figure}

Fig.~\ref{fig:indiv_IPR} shows the \ac{DOS} and \(tau\), the scaling exponent of the IPR with system size, for a representative set of parameters covering all three phases. The DOS is symmetric about \(0\) because of the particle hole symmetry of the model. At high temperatures, all of the eigenstates are localised in both the Mott and Anderson phases (with \(\tau \leq 0.07\) for our system sizes). We also checked that the states are localised by direct inspection. Note that there are in-gap states for instance at \(\omega_0\), below the upper band which are localized and smoothly connected across the phase transition. 

In the CDW phases at \(U=2\) and \(U=5\), we find for the states within the gapped CDW bands, e.g. at \(\omega_1\), scaling exponents \(\tau = 0.30\pm0.03\) and \(\tau = 0.15\pm0.05\), respectively. This surprising finding suggests that the CDW bands are partially delocalised with multi-fractal behaviour of the wavefunctions~\cite{eversAndersonTransitions2008a}. This phenomenon would be unexpected in a 1D model as they generally do not support delocalisation in the presence of disorder except as the result of correlations in the emergent disorder potential~\cite{croyAndersonLocalization1D2011,goldshteinPurePointSpectrum1977a}. However, we later show by comparison to an uncorrelated Anderson model that these nonzero exponents are a finite size effect and the states are localised with a finite \(\xi\) similar to the system size. As a result, the IPR does not scale correctly until the system size has grown much larger than \(\xi\). Fig.~[\ref{fig:indiv_IPR_disorder}] shows that the scaling of the IPR in the CDW phase does flatten out eventually. 

Next, we use the \ac{DOS} and the scaling exponent \(\tau\) to explore the localisation properties over the energy-temperature plane in Fig.~\ref{fig:band_opening}. Gapped areas are shown in white, which highlights the distinction between the gapped Mott phase and the ungapped Anderson phase. In-gap states appear just below the critical point, smoothly filling the bandgap in the Anderson phase and forming islands in the Mott phase. As in the finite~\cite{zondaGaplessRegimeCharge2019} and infinite dimensional~\cite{hassanSpectralPropertiesChargedensitywave2007} cases, the in-gap states merge and are pushed to lower energy for decreasing U as the \(T=0\) CDW gap closes. 
Intuitively, the presence of in-gap states can be understood as a result of domain wall fluctuations away from the AFM ordered background. These domain walls act as local potentials for impurity-like bound states~\cite{zondaGaplessRegimeCharge2019}.

In order to understand the localization properties we can compare the behaviour of our model with that of a simpler Anderson disorder model (DM) in which the spins are replaced by a \ac{CDW} background with uncorrelated binary defect potentials, see Appendix~\ref{app:disorder_model}. Fig.~[\ref{fig:indiv_IPR_disorder}] compares the FK model to the disorder model at different system sizes, matching the defect densities of the disorder model to the FK model at \(N = 270\) above and below the CDW transition. We find very good, even quantitative, agreement between the FK and disorder models, which suggests that correlations in the spin sector do not play a significant role. As we can sample directly from the disorder model, rather than through MCMC, the samples are uncorrelated. Hence we can evaluate much larger system sizes with the disorder model which enables us to pin down the correct localisation effects. In particular, what appear to be delocalized states for small system sizes eventually turn out to be states with large localization length. The localization length diverges towards the ordered zero temperature CDW state. Overall, we see that the interplay of interactions, here manifest as a peculiar binary potential, and localization can be very intricate and the added advantage of our 1D model is that we can explore very large system sizes for a complete understanding.  

\section{Discussion \& Conclusion}
The \ac{FK} model is one of the simplest non-trivial models of interacting fermions. We studied its thermodynamic and localisation properties brought down in dimensionality to 1D by adding a novel long-ranged coupling designed to stabilise the \ac{CDW} phase present in dimension two and above. Our hybrid \ac{MCMC} approach elucidates a disorder-free localization mechanism within our translationally invariant system. Further, we demonstrate a significant speedup over the naive method. We show that our long-range \ac{FK} in 1D retains much of the rich phase diagram of its higher dimensional cousins. Careful scaling analysis indicates that all the single particle eigenstates eventually localise at nonzero temperature albeit only for very large system sizes of several thousand.

Our work raises a number of interesting questions for future research. 
A straightforward but numerically challenging problem is to pin down the model's behaviour closer to the critical point where correlations in the spin sector would become significant. Would this modify the localisation behaviour? Similar to other soluble models of disorder-free localisation, we expect intriguing out-of equilibrium physics, for example slow entanglement dynamics akin to more generic interacting systems~\cite{hartLogarithmicEntanglementGrowth2020}. In a broader context, we envisage that long-range interactions can also be used to gain a deeper understanding of the temperature evolution of topological phases. One example would be a long-ranged \ac{FK} version of the celebrated Su-Schrieffer-Heeger model where one could explore the interplay of topological bound states and thermal domain wall defects. Finally, the rich physics of our model should be realizable in systems with long-range interactions, such as trapped ion quantum simulators, where one can also explore the fully interacting regime with a dynamical background field.  

\section{Acknowledgments}
We wish to acknowledge the support of Alexander Belcik who was involved with the initial stages of the project. We thank Angus MacKinnon for helpful discussions, Sophie Nadel for input when preparing the figures and acknowledge support from the Imperial-TUM flagship partnership. This work was supported in part by the Engineering and Physical Sciences Research Council (EPSRC)  \href{https://gtr.ukri.org/project/145404DD-ABAD-4CFB-A2D8-152A6AFCCEB7#/tabOverview}{Project No.~2120140}.

\appendix

\section{\label{app:balance}
DETAILED BALANCE
}
Given a \ac{MCMC} algorithm with target distribution \(\pi(a)\) and transition function \(\T\) the detailed balance condition is sufficient (along with some technical constraints \cite{wolffMonteCarloErrors2004}) to guarantee that in the long time limit the algorithm produces samples from \(\pi\). 
\[\pi(a)\T(a \to b) = \pi(b)\T(b \to a)\]

In pseudo-code, our two step method corresponds to two nested comparisons with the majority of the work only occurring if the first test passes:

\begin{lstlisting}[language=Python]
current_state = initial_state

for i in range(N_steps):
  new_state = proposal(current_state)

  c_dE = classical_energy_change(
                               current_state,
                               new_state)
  if uniform(0,1) < exp(-beta * c_dE):
    q_dF = quantum_free_energy_change(
                                current_state,
                                new_state)
    if uniform(0,1) < exp(- beta * q_dF):
      current_state = new_state

    states[i] = current_state
\end{lstlisting}

Defining \(r_c = e^{-\beta H_c}\) and \(r_q = e^{-\beta F_q}\) our target distribution is \(\pi(a) = r_c r_q\). This method has \(\T(a\to b) = q(a\to b)\A(a \to b)\) with symmetric \(p(a \to b) = \p(b \to a)\) and \(\A = \min\left(1, r_c\right) \min\left(1, r_q\right)\) 

 Substituting this into the detailed balance equation gives: 
\[\T(a \to b)/\T(b \to a) = \pi(b)/\pi(a) = r_c r_q\]

Taking the LHS and substituting in our transition function:
\begin{align}
\T(a \to b)/\T(b \to a) = \frac{\min\left(1, r_c\right) \min\left(1, r_q\right)}{ \min\left(1, 1/r_c\right) \min\left(1, 1/r_q\right)}
\end{align}

which simplifies to \(r_c r_q\) as \(\min(1,r)/\min(1,1/r) = r\) for \(r > 0\). 

\section{\label{app:disorder_model}
UNCORRELATED DISORDER MODEL
}

The disorder model referred to in the main text is defined by replacing the spin degree of freedom in the FK model \(S_i = \pm \tfrac{1}{2}\) with a disorder potential \(d_i = \pm \tfrac{1}{2}\) controlled by a defect density \(\rho\) such that \(d_i = -\tfrac{1}{2}\) with probability \(\rho/2\) and \(d_i = \tfrac{1}{2}\) otherwise. \(\rho/2\) is used rather than \(\rho\) so that the disorder potential takes on the zero temperature CDW ground state at \(\rho = 0\) and becomes a random choice over spin states at \(\rho = 1\) i.e the infinite temperature limit.
~
\begin{align}
H_{\mathrm{DM}} = & \;U \sum_{i} (-1)^i \; d_i \;(c^\dag_{i}c_{i} - \tfrac{1}{2}) \\
& -\;t \sum_{i} c^\dag_{i}c_{i+1} + c^\dag_{i+1}c_{i} \nonumber
\end{align}


\bibliographystyle{apsrev4-2}
\bibliography{main}

\end{document}